# Kinetics of Li transport in vanadium-based disordered rocksalt structures

Zinab Jadidi[1,2], Tina Chen[1,2], Luis Barroso-Luque[1,2], Gerbrand Ceder[1,2]

[1] *Department of Materials Science and Engineering, Berkeley CA 94720*
[2] *Materials Sciences Division, Lawrence Berkeley National Laboratory, Berkeley CA 94720*

*Abstract* - **Disordered rocksalt Li-excess (DRX) compounds have emerged as promising new cathode materials for lithium-ion batteries, as they can consist solely of resource-abundant metals and eliminate the need for cobalt or nickel. A deeper understanding of the lithium-ion transport kinetics in DRX compounds is essential for enhancing their rate performance. This study employs first-principles calculations, cluster expansion techniques, and kinetic Monte Carlo simulations to investigate the Li$^+$ transport properties in DRX Li$_{2-x}$VO$_3$, where $0 \leq x \leq 1$. Our findings underscore (i) the necessity of accounting for both tetrahedral and octahedral Li occupancy when predicting the transport properties in DRX materials, (ii) the factors influencing the variation in the diffusion coefficients with Li content in Li$_{2-x}$VO$_3$, and (iii) the impact of Li$^+$ correlated motion on the kinetics of Li$^+$ transport. We reveal that the relative stability of tetrahedral and octahedral Li determines the number of active sites within the percolation network, subsequently affecting the Li$^+$ transport properties. Furthermore, we demonstrate that the wide site-energy distribution causes correlated motion in Li$_{2-x}$VO$_3$, which hinders Li$^+$ transport. Although our study focuses on Li$_{2-x}$VO$_3$ as a model system, the insights gained apply to all DRX materials, given their inherently broader site-energy distributions.**

## INTRODUCTION

Disordered rocksalt cathodes were initially considered electrochemically inactive (1,2) due to the slow kinetics of Li$^+$ transport, resulting in subpar electrochemical performance. (3) However, the development of the so-called zero-transition-metal (0-TM) percolation theory facilitated the design of disordered rocksalt materials with improved Li$^+$ transport properties. (4,5) The 0-TM percolation theory (4,5) was developed based on the concept of octahedral Li occupancy and the tetrahedral site hop (TSH) diffusion mechanism. (6) In this mechanism, a Li$^+$ migrates from an octahedral site to an adjacent (edge-sharing) octahedral site via an intermediate tetrahedral site without any TM in the neighboring face-sharing octahedral site (0-TM channel). (4,5) According to the 0-TM percolation theory, Li$^+$ in a disordered structure can percolate if there is sufficient Li (an excess beyond the 1:1:2 Li:TM:O ratio of typical ordered rocksalt cathodes) to form low-barrier percolating 0-TM diffusion networks. (4,5) As disordered rocksalt structures with Li-excess (DRX) materials can be designed with a broader range of elements (4,7–16), the insights gained from the 0-TM percolation theory have provided valuable guidelines for the battery community. This has led to the development of cobalt-free and nickel-free DRX cathode materials capable of delivering specific energies up to 1000 Wh/kg. (3,7)

Although the 0-TM percolation theory offers valuable insights into the structural features that can influence Li$^+$ transport properties, it does not include all the factors affecting transport. As a result, its predictions do not always lead to high Li diffusivity (11,12,17) or the desired rate performance. (18) For instance, Li$_2$VO$_2$F exhibits considerable Li-excess but it is reported to have capacity fade due to seemingly slow Li$^+$ transport kinetics. (18)

A significant aspect not addressed in the current 0-TM percolation theory is the possibility of Li occupying tetrahedral sites in DRX materials (19–21), particularly in highly delithiated states where non-face-sharing tetrahedral Li sites can form. Consequently, alternative Li$^+$ hopping mechanisms, such as hops from a tetrahedral site to another tetrahedral site through an octahedral site or from an octahedral/tetrahedral site to a tetrahedral/octahedral site (20,21), are not considered in the 0-TM percolation analysis of DRX materials.

Another crucial aspect not included in the current percolation theory is the variability of the site energy which may lead to highly correlated Li$^+$ motion, and hereby significantly hinder transport. (6,22,23) For instance, even if a network of low-barrier pathways exists, the Li$^+$ motion may be correlated in such a way that it moves in circles or hops back and forth between the same sites, resulting in no net transport despite locally low migration barriers. (22,23) This type of highly correlated Li jumps have been shown to impede Li$^+$ transport in spinel Li$_x$TiS$_2$ (22) despite its fully connected 0-TM network.

To expand our understanding of the critical factors influencing Li$^+$ percolation in DRX materials, we develop a kinetic model that accounts for both tetrahedral and octahedral Li occupancy and the variable site energies of Li due to the diversity of local environments. In this paper, we study Li$_{2-x}$VO$_3$, where $0 \leq x \leq 1$, as a model system that has been investigated as a potential cathode material (24). We model this system using a lattice cluster expansion, fit to density functional theory (DFT) first-principles calculations, that includes quaternary disorder ($Li^+$/$V^{4+}$/$V^{5+}$ and vacancies) on octahedral sites and binary disorder ($Li^+$ and vacancies) on tetrahedral sites. The kinetics of $Li^+$ transport is modeled with this cluster expansion using the kinetic Monte Carlo method, employing the rejection-free (n-fold) algorithm. (25) Our findings reveal that the relative stability of the tetrahedral sites compared to the octahedral sites plays a crucial role in determining how $Li^+$ diffusivity varies with Li content. Furthermore, we demonstrate that the wide site energy distribution is the primary factor behind the correlated motion in $Li_{2-x}VO_3$. Our analysis recognizes a flat energy landscape as a critical design criterion for improved transport in DRX materials.

## METHODS

### A. Cluster expansion model

The cluster expansion (CE) method provides a formal representation of the energy landscape with respect to the configuration of species located on a predefined topology of lattice sites. (26–28) Extensive configurational sampling can be obtained by using Monte Carlo with the CE method, which is not possible directly using DFT. The CE model used in this work was constructed using the Statistical Mechanics on Lattices (smol) Python package (29) and was extensively described in our previous work on the cationic short-range ordering of $Li_{2-x}VO_3$, $0 \leq x \leq 1$. (30) We present a brief overview of our model and refer readers to the aforementioned references for further details.

The CE for the configurational energy is written in the form shown in equation 1:

$$E(\boldsymbol{\sigma}) = \sum_{\beta} V_{\beta} \langle \Phi_{\alpha}(\boldsymbol{\sigma}) \rangle_{\beta}. \quad (1)$$

The occupancy string $\boldsymbol{\sigma}$, in which each element $\sigma_i$ denotes an occupation variable for site i, is used to represent configurations. The summation in equation 1 encompasses all symmetrically distinct clusters of basis functions β. We use sinusoidal basis functions as detailed by Van de Walle et al. (31) The coefficients $V_{\beta}$ are the effective cluster interactions (ECI) weighted by the crystallographic multiplicity of the corresponding clusters. (26,32) The functions $\Phi_{\alpha}$ are cluster functions. (31)

Our CE model considers pair interactions up to 7.1 Å and triplet interactions up to 4 Å, based on a cubic cell with a lattice constant $a$ = 3.0 Å. An Ewald summation term is also included in the energy model to increase the accuracy and sampling stability of our CE by accounting for long-range electrostatic interactions. (15,33–35)

The degrees of freedom in the lattice model for $Li_{2-x}VO_3$, as illustrated in Figure 1, include {$Li^+$, $V^{4+}$, $V^{5+}$, vacancy} on the octahedral sublattice (24) and {$Li^+$, vacancy} on the tetrahedral sublattice. (19,21) The anion sites are fully occupied by $O^{2-}$ and hence do not need to be represented in the CE. (24) Sparse group lasso regularization was used to fit the ECIs of this high-component system using DFT energies of 450 symmetrically distinct configurations. In 10-fold cross-validation the resulting fit reproduced DFT with a root-mean-squared error (RMSE) of 13.1 meV per site and in-sample RMSE of 12.4 meV per site.

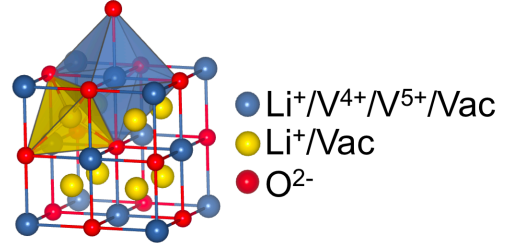

**FIG. 1.** Rocksalt lattice of $Li_{2-x}VO_3$, $0 \leq x \leq 1$. The blue spheres correspond to the face-centered cubic (FCC) lattice of octahedral sites, which can be occupied by $Li^+$, $V^{4+}$, $V^{5+}$, or a vacancy. The yellow spheres correspond to tetrahedral sites, which can be occupied by $Li^+$ or a vacancy. The red spheres form the anion lattice, which is fully occupied by oxygen and does not need to be included in the degrees of freedom of the CE.

### B. First-principles DFT calculations

The structure–energy dataset used to fit the CE was obtained using DFT (36,37) implemented in the Vienna ab-initio simulation package (VASP) (38,39) with the projector-augmented wave (PAW) method. (40,41) The PBE exchange-correlation functional (42) with rotationally averaged Hubbard (U = 3.1 eV) correction was used based on a previously reported calibration to oxide formation energies (3.1 eV for V). (15,43) A plane-wave cutoff of 520 eV, reciprocal-space discretization of 25 K-points per Å, electronic convergence of $10^{-5}$ eV and 0.01 eV Å$^{-1}$ for forces were used in all of our calculations.

The Nudged Elastic Band (NEB) calculations are conducted using five images along the diffusion paths for single-vacancy and divacancy tetrahedral site hop (TSH) mechanisms at compositions of $LiVO_3$, $Li_{1.5}VO_3$, and $Li_2VO_3$ in supercells of 3×3×3 or 3×4×3, containing 27 or 36 oxygen atoms. Configurations for the NEB calculations are created by Monte Carlo at temperatures of 300 K and 5000 K to generate various local environments for $Li^+$ migration. In the NEB calculations, it is ensured that the distance between the migrating $Li^+$ and its periodic image exceeds 8 Å in order to prevent any interaction between periodic images. All the NEB calculations are performed using the GGA method without the Hubbard-U correction to prevent potential convergence issues arising from electron localization at different atomic centers along the migration path. (44,45)

### C. Transport properties

In solid crystalline intercalation compounds, such as DRX materials, the diffusion of $Li^+$ typically occurs through a substitutional diffusion mechanism, in which $Li^+$ migrates to adjacent vacant sites. (6,22,46–49) The presence of diverse local environments in DRX materials results in a dependency of the $Li^+$ migration energy barrier on its local surroundings. Kinetic Monte Carlo (KMC) simulations can combine a cluster expansion to incorporate the $Li^+$-$Li^+$ and $Li^+$-host energetics with a local cluster expansion to model the variations in migration energy. (6) We show below a simplified model to capture the dependence of migration energy on environment.

To derive the diffusion coefficients from the Li trajectories we employ the Green–Kubo approach, derived from the fluctuation theorem at equilibrium. (50–52) The chemical diffusion coefficient, $D_{Chemical}$, represents the relationship between the flux of migrating species, and their concentration gradients. (51) $D_{Chemical}$ can be calculated by multiplying the thermodynamic factor, Θ, with the jump diffusion coefficient, $D_{Jump}$ (51):

$$D_{Chemical} = \Theta D_{Jump}. \quad (2)$$

The thermodynamic factor Θ, given by equation 3, measures the deviation of the solution thermodynamics from the non-interacting ideal solution. When $Li^+$–$Li^+$ interactions are negligible (in the dilute-limit case), the thermodynamic factor is equal to 1.

$$\Theta = \frac{\partial\left(\frac{\mu}{k_B T}\right)}{\partial \ln(x)}. \quad (3)$$

In equation 3, μ represents the Li chemical potential, $k_B$ denotes Boltzmann's constant, T signifies the absolute temperature, and $x$ corresponds to the Li concentration. Because μ is the negative of the cathode voltage (53) the thermodynamic factor is related to the voltage profile. In this study, the thermodynamic factor is computed by scanning the Li chemical potential in steps of 0.05 eV using semi-grand canonical Monte Carlo (SGC MC) simulations at T = 300 K. The thermodynamic and transport analysis are conducted in a 9×8×9 supercell (comprising a 648-oxygen supercell) of $Li_{2-x}VO_3$ with Li/V metal configurations obtained by equilibrating at 1800 K. The selection of 1800 K was established in previous work and is based on the similarity of the simulated and experimental voltage profiles and XRD patterns at this temperature. (30)

The jump diffusion coefficient $D_{Jump}$ in equation 2 measures the fluctuation of the center of mass of all diffusing Li-ions and can be determined as (54,55):

$$D_{Jump} = \frac{1}{2dt}\frac{1}{N}\langle\left[\sum_{i=1}^{N}\Delta\vec{R_i}(t)\right]^2\rangle. \quad (4)$$

In equation 4, $d$ represents the dimension of the lattice on which diffusion occurs (for DRX $Li_{2-x}VO_3$, $d$ is 3), $t$ denotes time, N refers to the number of diffusing Li ions, and $\Delta\vec{R}_i(t)$ is the displacement vector of the ith ion after time t. The angular brackets signify the ensemble average after time $t$.

Compared to $D_{Jump}$, the tracer diffusion coefficient, $D_{Tracer}$, is based on the fluctuation of the position of individual Li-ions and can be determined as follows:

$$D_{Tracer} = \frac{1}{2dt}\frac{1}{N}\sum_{i=1}^{N}\langle[\Delta\vec{R}_i(t)]^2\rangle. \quad (5)$$

The tracer diffusion includes self-correlation caused by the diffusion topology and the energy landscape. The deviation of $D_{Tracer}$ from the random-walk scenario is measured by the correlation factor: (22)

$$f = \frac{\langle\vec{R}(t)^2\rangle}{na^2}. \quad (6)$$

In equation 6, $\langle\vec{R}(t)^2\rangle$ represents the mean squared displacement of a Li ion after a time t. The variable $n$ refers to the number of jumps $Li^+$ has executed during the simulation period, and $a$ is the hop distance.

The rate constants $\Gamma_{ij}$ for $Li^+$ hopping from site i to site j can be determined from transition state theory (56):

$$\Gamma_{ij} = \nu^* exp\left(-\frac{\Delta E_{ij}}{K_B T}\right). \quad (7)$$

where $\nu^*$ is the vibrational prefactor, estimated to be $10^{13}$. (6) The migration activation barrier for $Li^+$ hopping from site i to site j is represented by $\Delta E_{ij}$. As $\Delta E_{ij}$ depends on the local environment, it is crucial to parameterize its value in accordance with the variations in these local environments. (6,47,49)

When tetrahedral Li sites can be occupied, more elementary diffusion steps are possible than when Li only occupies octahedral sites. (21) To simplify our kinetic model, we allow $Li^+$ to hop between octahedral sites and their nearest-neighbor (face-sharing) tetrahedral sites and vice versa, as most of the diffusion mechanisms in DRX materials (21) can be broken down into these two fundamental hops. Hence, we assign a value of 1.82 Å to parameter $a$ in equation 6, as it corresponds to the nearest-neighbor distance between octahedral and tetrahedral sites. $Li^+$ migration in DRX compounds can also potentially occur directly from a tetrahedral site to an edge-sharing tetrahedral site. (21) However, we have excluded this mechanism for two reasons. Firstly, the barrier for $Li^+$ migration through edge-sharing tetrahedral-tetrahedral hops has been found to reach up to 900 meV. (21) Secondly, including this mechanism will not significantly impact our simulation's final results since most low-barrier $Li^+$ migration mechanisms predominantly occur through the triangular face rather than the edge. (21)

The energy of a $Li^+$ ion in a tetrahedral site depends strongly on the occupancy of its four face-sharing octahedral sites. In our kinetic model, we account for all $Li^+$ transitions from octahedral to tetrahedral sites that do not face-share with any vanadium-occupied sites, or face-share with a single vanadium-occupied site and two vacant sites. These transitions are abbreviated as 0-TM and 1-$TM_{di-vac}$ in line with notation in earlier papers on diffusion in rocksalt-like materials. (4,5) Due to the high migration energy observed in the 1-$TM_{single-vac}$ hop (Figure S1), we omitted it from our KMC simulations. Hops into tetrahedral sites with two face-sharing vanadium have even higher energy and can be excluded from our simulation.(4) Migration barriers for face-sharing hops are obtained from NEB calculations of the TSH mechanism when the tetrahedral site represents a local minimum in the minimum energy pathway. To integrate migration barriers with the site energy landscape of the cluster expansion, we use the kinetically resolved activation (KRA) barrier in KMC simulations as originally proposed by Van der Ven. (6) The KRA is calculated by subtracting the average energy of the hop's endpoints from the energy of the activated state. (6) In our KMC simulations, we employed a single KRA value for all hops, and we provide the rationale for this decision in Figure 2.

KMC simulations are conducted at 300 K using the rejection-free algorithm. (57,58) The initial state of each KMC simulation is done in a 9×8×9 supercell for 12 compositions of $Li_{2-x}VO_3$. At each KMC step (i.e., a $Li^+$ hop to a nearby vacancy), we randomly select a migrating $Li^+$ and choose a jump based on a probability proportional to its rate. (25) We perform 50,000 KMC steps starting from each initial configuration and average the $Li^+$ trajectories after 20,000 steps to ensure the system is fully equilibrated.

## RESULTS

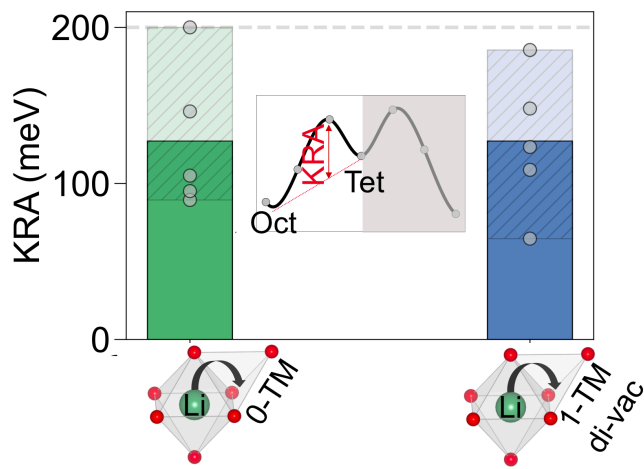

**FIG. 2.** The KRA barriers for $Li^+$ hops from an octahedral site to 0-TM (green) and 1-$TM_{di\text{-}vac}$ tetrahedral sites (blue) are displayed. The solid bars represent the average value of the KRAs, and the hatched bars illustrate the range between the maximum and minimum values of the KRAs for each hop category. These KRAs were determined from 10 NEB calculations of TSH mechanisms on different Li/V configurations. In these calculations, the intermediate tetrahedral site acts as a local minimum for the migrating $Li^+$ (as shown in the inset plot). The circle markers indicate the data points for each hop category. The gray dotted line represents the upper limit of the KRA values, set at 200 meV, which is used as input for all octahedral-to-tetrahedral (or vice versa) hops in the KMC simulation.

Figure 2 displays the KRA values for $Li^+$ hops from octahedral sites into a tetrahedral site face-sharing with different number of TM's and vacancies. Values are plotted for a tetrahedron with no TM face-sharing and either 1, 2 or 3 vacancies (green) and for the single -TM face-sharing with 2 vacancies (blue). The minimum energy pathways for all 10 NEB calculations used to determine the KRA values are presented in Figure S2.

The KRA values for both 0-TM and 1-$TM_{di\text{-}vac}$ hops range from ~ 70 to 200 meV, with their mean values centered at ~130 meV. The fact that the migration barrier of some 1-$TM_{di\text{-}vac}$ hop is comparable to that of 0-TM hops may be related to the ability of the face-sharing vanadium to relax away from the center of its octahedron. (20) This effect, especially pronounced for ions with $d^0$ electron configuration like $V^{5+}$ (59,60), reduces the electrostatic repulsion between $Li^+$ in the tetrahedral site and the face-sharing V. This relatively low energy of tetrahedral Li–V face-sharing (20) is consistent with these features being observed in the equilibrium modeling of both $Li_{2-x}VO_3$ (30) and $Li_{3+x}V_2O_5$ DRX materials. (20) As there is no significant difference between 0-TM and 1-$TM_{di\text{-}vac}$ hops, we used a single KRA value for the KMC simulation to further simplify the kinetic model. The energy difference between the initial and final states of the hop, captured by the CE, accounts for the difference between the migration barriers for 0-TM and 1-$TM_{di\text{-}vac}$ hops. Using a single KRA value for parametrizing the energy barrier of $Li^+$ hops has been successfully employed in other transport studies based on KMC simulations. (23) The selected KRA value for any octahedral-to-tetrahedral (or vice versa) hop is set at 200 meV, equal to the upper limit of the KRA values presented in Figure 2. Utilizing a single KRA value can influence the absolute magnitude of $Li^+$ diffusivity, which we further elaborate on in the discussion section.

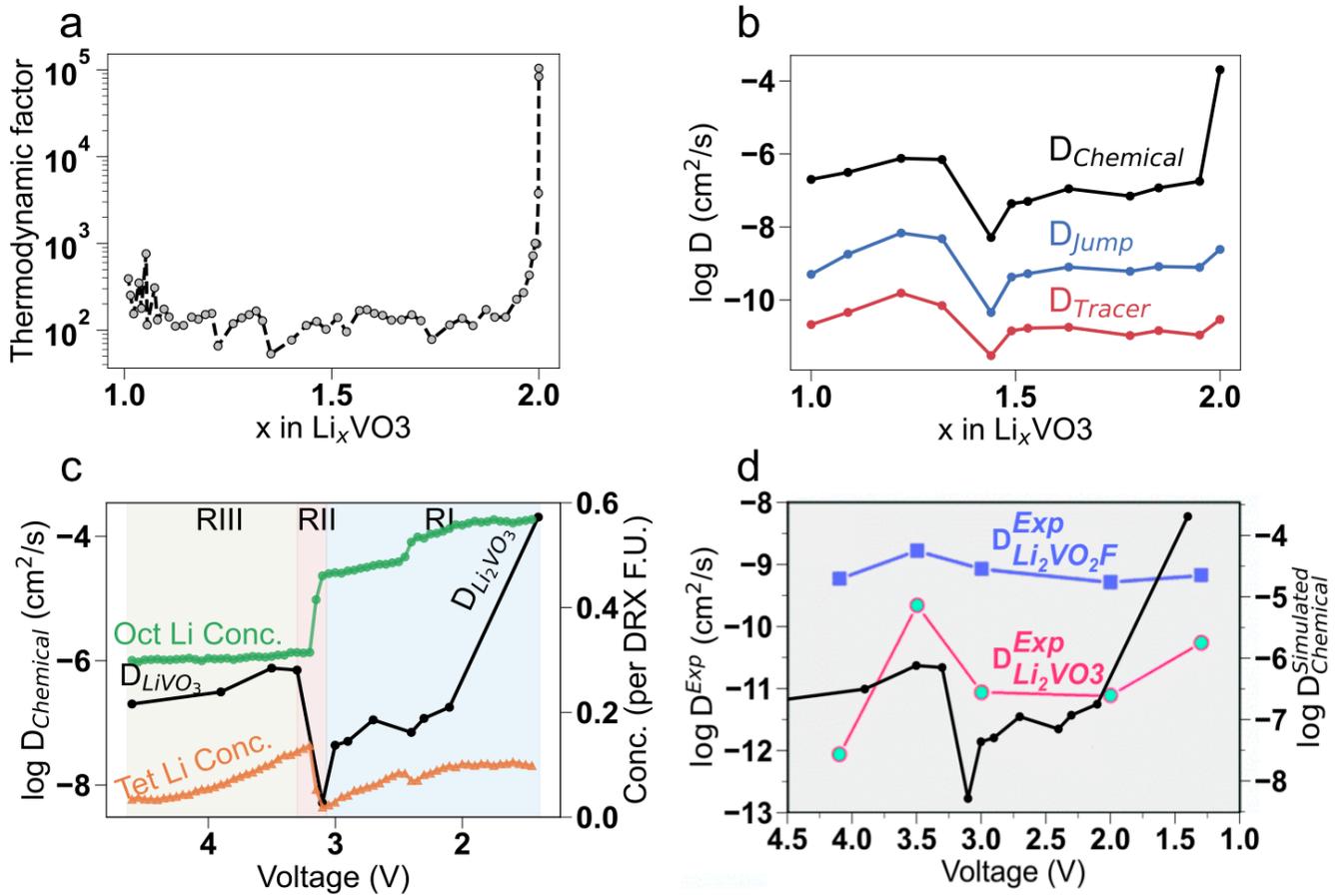

**FIG. 3. (a)** The calculated thermodynamic factor, $\Theta$, is shown as a function of Li content, x, in $Li_xVO_3$, at 300 K. The thermodynamic factor diverges at $x_{Li}$ = 2, corresponding to the fully lithiated composition. **(b)** The calculated $D_{Tracer}$ (red), $D_{Jump}$ (blue), and $D_{Chemical}$ (black) are displayed as a function of the Li content in $Li_xVO_3$ at 300 K, in units of $cm^2/s$. **(c)** The computed concentrations of octahedral Li (green) (30), tetrahedral Li (orange) (30), and $D_{Chemical}$ (black) as a function of voltage (30). The regions RI, RII, and RIII correspond to the voltage ranges of (1.4 to 3.1), (3.1 to 3.3), and (3.3 to 4.6), respectively. **(d)** Experimental diffusivity of $Li_2VO_3$ (red curve) and $Li_2VO_2F$ (blue curve) obtained by Chen et al. (24) and simulated diffusivity of $Li_2VO_3$ (black curve) calculated by KMC simulations.

Figure 3a displays the thermodynamic factor calculated at 300 K as a function of Li content in $Li_xVO_3$. The value of the thermodynamic factor is on the order of $10^2$ but diverges to much larger values for the fully lithiated composition as the thermodynamic factor is proportional to the inverse of the vacancy concentration. (6,22,47) Such behavior of the thermodynamic factor has been previously observed in certain layered and spinel systems. (6,22,47)

Figure 3b displays $D_{Chemical}$ (black), $D_{Jump}$ (blue), and $D_{Tracer}$ (red) calculated using equations 2, 4, and 5 as a function of Li content in $Li_xVO_3$. The magnitude of $D_{Tracer}$ ranges from $10^{-11}$ to $10^{-10}$ $cm^2/s$, whereas that of $D_{Jump}$ spans from $10^{-10}$ to $10^{-8}$ $cm^2/s$. In contrast, the magnitude of $D_{Chemical}$ mainly falls between $10^{-8}$ and $10^{-6}$ $cm^2/s$, except at the fully lithiated state, where it is on the order of $10^{-4}$ $cm^2/s$. Consistent with other kinetic studies on certain layered and spinel systems, $D_{Tracer}$ is the lowest among the diffusion coefficients. (6,22,47) This finding occurs because $D_{Jump}$ and $D_{Chemical}$ rely on the collective displacements of $Li^+$ rather than the average displacements of individual ions, which is the case for $D_{Tracer}$.

A noticeable jump in all the simulated diffusion coefficients is observed, with the lowest point occurring at the $Li_{1.44}VO_3$ composition and the highest point at $Li_{1.22}VO_3$. To investigate the reason for this jump, we present the concentrations of octahedral Li (green) (30) and tetrahedral Li (orange) (30) and the corresponding $D_{Chemical}$ as a function of voltage (30) in Figure 3c. In order to directly compare the simulated diffusivity with the experimental diffusivity of $Li_2VO_3$ (red curve in Figure 3d) obtained by Chen et al. (24), we analyze these quantities as a function of voltage instead of Li concentration in both Figure 3c and 3d.

Figure 3c highlights three regions. Region I (RI), which covers the voltage range of (1.4 to 3.1), corresponds to the delithiation of $Li_2VO_3$ down to $Li_{1.49}VO_3$ (as shown in Figure 3b). In this region, the magnitude of $D_{Chemical}$ initially decreases from $Li_2VO_3$ to $Li_{1.95}VO_3$ and then remains approximately constant at $\sim 10^{-7}$ $cm^2/s$. The experimental diffusivity of this material exhibits a similar change as a function of voltage within this voltage range, as shown in Figure 3d. (24) Additionally, we observe that both the tetrahedral and octahedral Li concentrations decrease in this voltage range.

Region II (RII), covering the voltage range of (3.1 to 3.3), corresponds to the delithiation of $Li_{1.49}VO_3$ to $Li_{1.32}VO_3$. As depicted in Figure 3c, we observe an increase in diffusivity along with the concentration of tetrahedral Li while the concentration of octahedral Li decreases. The experimental diffusivity of this material exhibits a similar jump within the similar voltage window. (24) The simultaneous decrease/increase in the octahedral/tetrahedral Li concentration suggests that some of the octahedral Li migrates to the tetrahedral sites. (30) This migration has also been observed for the fluorinated iso-structure of this material at this composition, $Li_{1.44}VO_2F$. (61) Such collective migration phenomena are known to occur in ordered spinel compounds (62); however, they have not yet been well characterized for cation-disordered materials. This correlation between the magnitude of diffusivity as a function of Li content and the concentrations of Li in tetrahedral and octahedral sites underscores the importance of considering both tetrahedral and octahedral sites in the transport analysis of DRX materials.

Region III (RIII) spans the voltage range of (3.3 to 4.6) and corresponds to the delithiation of $Li_{1.32}VO_3$ to $LiVO_3$. In this region, we observe a decrease in diffusivity, which aligns with the behavior of the experimental diffusivity of $Li_xVO_3$ as a function of voltage. (24) In RIII, the Li concentration in octahedral sites remains constant while the tetrahedral Li concentration decreases. This suggests that Li extraction from $Li_{1.44}VO_3$ primarily occurs from high-voltage tetrahedral sites (30), contributing to the lower diffusivity in this range. Notably, the experimental diffusivity of this material decreases more rapidly (24) than the simulated diffusivity in RIII. In the discussion section, we use our understanding derived from the $Li_2VO_3$ compound to help interpret the experimental diffusion behaviors of $Li_2VO_2F$.

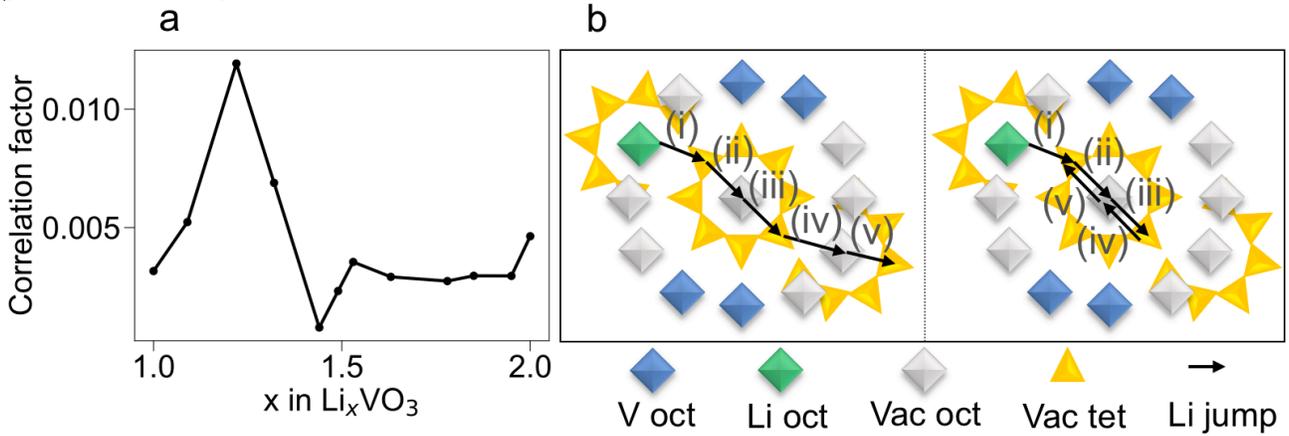

**FIG. 4.** (a) The calculated correlation factor as a function of Li content, x, in $Li_xVO_3$. (b) A two-dimensional schematic of a Li-ion migrating in a rocksalt structure is shown. The left panel represents the case where the correlated motion of Li-ions is not considered, whereas the right panel represents the case where the correlated motion is included. In a rocksalt structure, each octahedral site (shown with diamonds) edge-shares with 12 other octahedral sites and face-shares with 8 tetrahedral sites (shown with triangles). The V-ions are represented by blue diamonds, while the green diamonds, gray diamonds, and yellow triangles represent octahedral Li sites, octahedral vacant sites, and tetrahedral vacant sites, respectively. The black arrows denote the $Li^+$ jumps, which are labeled sequentially from (i) to (v).

Figure 4a displays the calculated correlation factor plotted against the Li content in $Li_xVO_3$. The correlation factor ranges between $10^{-3}$ and $10^{-2}$ with the highest value at the composition $Li_{1.22}VO_3$ and the lowest value at $Li_{1.44}VO_3$. The observed correlation factor in $Li_xVO_3$ is consistent with the predicted range of $10^{-3}$ to $10^{-2}$ in the DRX transport model developed by Anand et al. (63) For comparison, the correlation factor for diffusion in layered $LiCoO_2$ (6) and the spinels $Li_{1+x}Ti_2O_4$ (47) and $Li_xTiS_2$ (22) is on the order of $10^{-1}$ across varying Li content, while the predicted correlation factor for $Mg_xTiS_2$ (23) can be as low as $10^{-5}$. Unlike $Li_xCoO_2$ (6) and the spinels $Mg_xTiS_2$ (23), $Na_xTiS_2$ (23), $Li_xTiS_2$ (23), which show an increase in the correlation factor with delithiation (i.e., increasing vacancy concentration), the correlation factor of DRX $Li_xVO_3$ does not show a consistent trend with increasing vacancy concentration. This finding indicates that factors other than the vacancy concentration play a role in determining $Li^+$ correlated motion in this DRX system.

To enhance our understanding of $Li^+$ correlated motion, Figure 4b depicts how Li-ions move in a local environment based on the results of the KMC simulation. The left panel of Figure 4b shows $Li^+$ migration from one octahedral site to another without significant self-correlation. However, in systems in which there is significant correlation, $Li^+$ hops back and forth between specific sites, slowing its progress to the following site, as depicted in the right panel of Figure 4b. This type of correlated motion is commonly observed and has been extensively studied in ionic conductors and electrolytes (64–66) and is known as *self-correlated* motion, reflected in the $D_{Tracer}$ value. (67)

It has been mathematically demonstrated that the connectivity of the sites impacts the correlated motion. (67) The connectivity of the sites can be disrupted if a high-barrier channel obstructs a network. In such cases, migrating $Li^+$ cannot move past the block or percolate through the structure.

(4,5) Both Li-excess and cation short-range ordering influence the formation of a percolating network in DRX. (11,33,68,69) As previously mentioned, Li-excess results in the formation of $Li_4$ clusters, whereas cation short-range ordering determines the extent of $Li_4$ tetrahedral connectivity in DRX materials. (11,33,68,69) Another factor responsible for the correlated motion is the ionic interactions that introduce non-uniformity in the energy landscape. (67) When an ion jumps uphill it is now more likely to jump backward to the site it came from. As stated by Anand et al. (63), when the standard deviation of the site-energy distribution exceeds $k_BT$, the fraction of Li sites involved in percolation can be significantly lower (63) than what the 0-TM percolation theory predicts. (4,5) It is worth noting that the model developed by Anand et al. represents Li–TM and Li–Li interactions through variations in site energies, considering all sites as Li/vacancy sites. (63)

energy differences that occur at composition $Li_{1.22}VO_3$ (green) the distribution narrows to 104 meV, while at composition $Li_{1.44}VO_3$ (orange) the distribution widens to 148 meV. Comparison of these hop distributions with the correlation factor in Figure 4a suggests that flatter energy landscapes, characterized by a narrower hop energy distribution, increase the correlation factor and thereby the effective diffusivity, as suggested by Anand. (63)

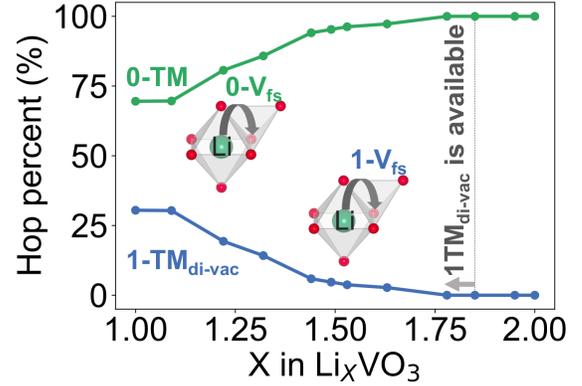

**FIG. 6.** The percentage of $Li^+$ hops from octahedral to tetrahedral sites that face-share with no vanadium (0-TM) (represented in green) or with one vanadium (depicted in blue) as function of Li composition.

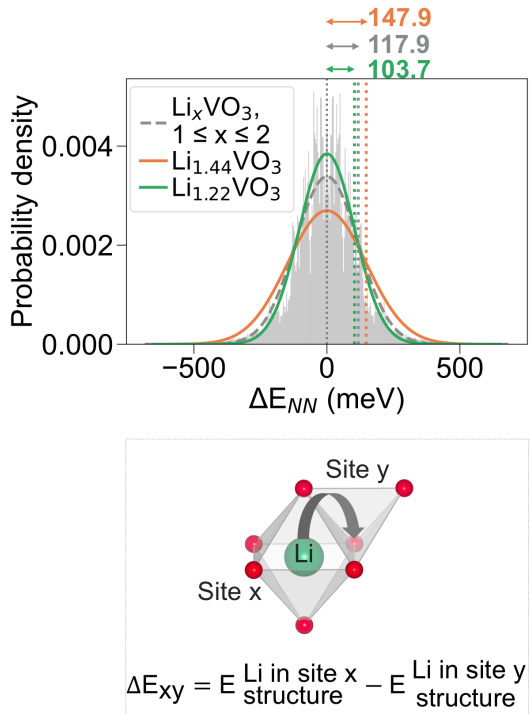

**FIG. 5.** The gray histogram in the above panel illustrates the energy difference between nearest-neighbor sites ($\Delta E_{NN}$) along the hops in the kinetic Monte Carlo simulation in the composition range of $Li_xVO_3$ ($1 \leq x \leq 2$). The dotted gray curve represents the fitted normal distribution to this histogram. The orange and green curves correspond to the normal distribution curves fitted to the energy differences between nearest-neighbor sites along the $Li^+$ hops restricted to compositions $Li_{1.44}VO_3$ and $Li_{1.22}VO_3$, respectively. The values 103.7, 117.9, and 147.9 at the top of the plot indicate the standard deviations of $\Delta E_{NN}$ for compositions $Li_{1.22}VO_3$, $Li_xVO_3$ ($1 \leq x \leq 2$), and $Li_{1.44}VO_3$, respectively. The schematic on the bottom further clarifies the meaning of $\Delta E_{NN}$ along the hop.

Figure 5 presents the energy difference between the nearest-neighbor sites along the $Li^+$ hop ($\Delta E_{NN}$). The gray histogram in the background of Figure 5 illustrates the distribution of $\Delta E_{NN}$ within the composition range of $Li_xVO_3$ ($1 \leq x \leq 2$). The dotted gray curve represents the normal distribution fitted to the $\Delta E_{NN}$ histogram and shows a standard deviation of approximately 118 meV. When fitted to only

The initial premise of the 0-TM percolation theory underscored the significance of incorporating 0-TM in transport analysis. (4,5) Our KMC model provides the opportunity to investigate the environment through which Li ions hop. We present in Figure 6 the percentage of hops from octahedral to 0-TM and 1-TM tetrahedral sites. Our results reveal a consistent predominance of $Li^+$ hops from octahedral to 0-TM tetrahedral sites over those to 1-$TM_{di\text{-}vac}$ tetrahedral sites, consistent with the basic idea of 0-TM percolation theory. (4,5) The lack of availability of 1-$TM_{di\text{-}vac}$ at high Li content is not unexpected, considering prior work on $LiCoO_2$ in which it was observed that the divacancy mechanism only becomes accessible for $x_{Li} < 0.8$. (6)

Our analysis proposes that as the number of vacancies increases towards $x_{Li} = 1$, hops to a tetrahedral site with face-sharing vanadium (1-$TM_{di\text{-}vac}$) occur, but remain minority pathways. Although our analysis has been centered on octahedral-to-tetrahedral hops rather than octahedral–tetrahedral–octahedral hops, our findings confirm that as a general rule 0-TM channels do provide the pathway for Li percolation though extensions to include the 1-$TM_{di\text{-}vac}$ mechanism could be relevant near the top of charge.

## DISCUSSION

In this paper we investigated the macroscopic diffusion mechanism in the disordered rocksalt $Li_xVO_3$ by combining a cluster expansion for the energy landscape and limited migration hops from octahedral to tetrahedral sites and vice versa. The incorporation of the tetrahedral site as a distinct lithium site enables us to better understand the role it plays in delithiation and its kinetics. Specifically, Figure 3c demonstrates the significant impact of the relative stability between Li in tetrahedral and octahedral sites on the variation of Li$^+$ diffusivity with Li content.

In contrast to certain layered and spinel systems (6,22,23,47), the diffusion coefficient (Figure 3b) and correlation factor (Figure 4a) of Li$_x$VO$_3$ do not improve with an increasing vacancy concentration. Moreover, empirical evidence shows that the diffusivity of DRX materials (11,12,17,24) remains relatively stable even when the Li concentration varies. This indicates the presence of other influential factors that overshadow the effect of vacancy concentration. The distribution of the Li site energy in DRX structures is greatly affected by the variation in the number and chemistry of the surrounding TM around the Li sites, which is determined by the strength of the Li–TM interaction. (19) For Li$^+$ to percolate efficiently, there must be a network of low-barrier local environments where all the nearest-neighbor sites are in a similar energy range. (63,66) Notably, the importance of the energy landscape in the kinetics of DRX materials is highlighted by Figure 5 and the recent publication by Anand et al. (63) For ordered layered structures such as LiCoO$_2$, where all the Li sites have the same number of TM around them, overcoming the energy difference between the initial and final states of the hop is not a significant issue, as the only difference is in the vacancy concentration. (6) However, Figure 5 reveals that for DRX Li$_x$VO$_3$, even the composition with highest diffusivity (Li$_{1.22}$VO$_3$) can exhibit a difference in nearest-neighbor site energies exceeding 400 meV. This substantial variation in nearest-neighbor site energy within Li$_x$VO$_3$ can help to explain the low correlation factor in Li$_x$VO$_3$ comparing to some layered and spinel systems. (6,22,23)

Our results reveal that the abrupt enhancement in both theoretical and experimental diffusivity shown in Figure 3d, (24) between the compositions of Li$_{1.22}$VO$_3$ (V = 3.5 V) and Li$_{1.44}$VO$_3$ (V = 3.1 V), can be attributed to the somewhat sudden transition of Li$^+$ from octahedral to tetrahedral sites, as demonstrated in Figure 3c. This same transition from octahedral to tetrahedral Li$^+$ migration has been computationally observed in Li$_{1.44}$VO$_2$F, a fluorinated isostructural counterpart of Li$_2$VO$_3$.(61) Interestingly, the experimental diffusivity of Li$_2$VO$_2$F (shown in Figure 3d) (24) exhibits a jump at 3.5 V (~ Li$_{1.44}$) (24), similar to the theoretical and experimental findings observed in the oxide compound, though with a less pronounced peak. (24) Collective migration phenomena such as these are recognized in ordered spinel compounds (62), yet their occurrence within DRX compounds is somewhat surprising. A potential explanation for this collective motion could be that following the removal of 28% of Li (from Li$_2$VO$_3$ to Li$_{1.44}$VO$_3$), the energy for Li to occupy tetrahedral and octahedral sites begins to converge due to the creation of vacancies on octahedral sites (which lower the energy of tetrahedral occupancy). Support for this explanation is provided in Figure 5, where the distribution of hop energies in the Li$_{1.22}$VO$_3$ composition exhibits the smallest standard deviation, indicating a dense network of available sites with comparable energies. Consequently, the correlation factor and the diffusion coefficient increase from Li$_{1.44}$VO$_3$, peaking at the composition Li$_{1.22}$VO$_3$.

An increase in the stability of Li$^+$ in tetrahedral sites can enhance the diffusivity by providing more available sites for Li$^+$; however, when the site energy for Li in tetrahedral sites becomes too low, their occupation by Li$^+$ can also impede migration through these sites, leading to lower experimental (24) and theoretical diffusivities as observed in highly delithiated states. Notably, the experimental diffusivity of Li$_x$VO$_3$ (24) declines more rapidly near the end of the charge than the theoretical diffusivity.

The experimental diffusivity of Li$_2$VO$_3$ has been reported to fall in the range of 10$^{-12}$ to 10$^{-10}$ cm$^2$/s (Figure 3d) (24), which is around four orders of magnitude lower than our simulated chemical diffusivity. Other kinetic studies have also observed this disparity between the experimental and computed diffusivity. (23,65) This discrepancy may have multiple origins. One of the primary problems when comparing experimental and simulated diffusivities is the challenge in estimating the active electrode and electrolyte surface area. (22) In the past, everything from the active electrode area to the total particle surface area has been used leading to very different diffusion constants. Assumptions in the theory models, such as the choice of single KRA and prefactor values, the diffusion mechanisms considered, may also contribute to the deviation between the experimental and theoretical diffusivity, though it seems unlikely that these would lead to variations that are more than an order of magnitude. The lack of any consideration of electronic transport limitations, either at the particle or electrode level may further complicate comparison between theory and experiment. Consequently, kinetic studies, including ours, have primarily concentrated on how the diffusivity varies with Li content rather than focusing on its absolute magnitude. (23)

The probability of finding tetrahedral sites in Li$_{2-x}$VO$_3$ face-sharing with 1-V, is twice as high as the probability of tetrahedral sites without face-sharing, as illustrated in Figure S3. Nevertheless, Figure 6 demonstrates that nearly all octahedral-to-tetrahedral hops within the composition range of Li$_{1.63}$VO$_3$ to Li$_2$VO$_3$ are into tetrahedral sites without V face-sharing. Beginning with the composition Li$_{1.32}$VO$_3$, we observe that over 10% of hops are to 1-TM$_{di-vac}$ tetrahedral sites, and this fraction rises to approximately 25% for LiVO$_3$. Still, our findings indicate that the substantial energy difference between octahedral and 1-TM tetrahedral sites makes the 0-TM network the most relevant pathway for Li$^+$ transport in DRX materials.

This study, along with the one conducted by Anand et al. (63), supports the notion that a flat energy landscape improves effective Li$^+$ percolation in DRX structures. However, no established strategy exists for creating a flat energy landscape in such structures. A study on the Li$_2$VO$_2$F compound demonstrated that substituting some of the V with Fe or Ti can enhance the electrochemical cyclability and reduce the voltage slope (70), suggesting a flatter energy landscape. Research on high-entropy disordered cathodes (71) has indicated that using a variety of TMs can decrease short-range order and promote long-range Li$^+$ transport. (71) Although introducing multiple TMs reduces short-range ordering (71), further investigation is needed to understand how this reduction influences the site energy distribution and whether it improves Li$^+$ transport properties by flattening the energy landscape or if other factors contribute to the enhancement.

The positive impact of F on Li$^+$ transport has also been demonstrated in other studies. (33) Comparing the experimental diffusivity of Li$_2$VO$_3$ and Li$_2$VO$_2$F (Figure 3d) (24) further emphasizes the advantages of F, as the fluorinated compound exhibits a diffusivity of approximately 1 to 3 orders of magnitude higher than that of the oxide compound. However, the amount of F must be carefully selected, as F can also immobilize Li due to strong Li–F bonds when a high concentration of vacancies is present. (69)

Due to the complexity of Li$^+$ transport in the DRX materials, accurately predicting Li$^+$ transport properties requires information on the site energy distribution and interactions, which can be obtained through advanced modeling techniques such as the CE and KMC simulations employed in this work. Our findings emphasize the importance of understanding the factors contributing to a flat energy landscape in DRX materials. More precise results will be achieved if the comparison of Li$^+$ percolation across different DRX materials is based on both 0-TM connectivity network as well as their energy landscape.

## CONCLUSION

In this study, we used KMC simulations to examine the kinetics of Li$^+$ transport in a Li$_2$VO$_3$ DRX material. Our findings underscore the necessity of considering the Li occupancy in both tetrahedral and octahedral sites to understand any potential collective ion migration and its possible impact on the macroscopic diffusivity. Our results show that the diffusivity and correlation factor of Li$_x$VO$_3$ do not exhibit enhancement as the vacancy concentration increases, which contrasts with the patterns observed in kinetic studies of some layered and spinel systems. This observation suggests that an uneven energy landscape within the DRX structures, can outweigh the influence of vacancy concentrations. Our findings continue to demonstrate that while 1-TM$_{di-vac}$ channels play a role in the diffusion process, it is the 0-TM channels that remain the primary channels, especially in highly lithiated states where the vacancy concentration is insufficient to activate the 1-TM$_{di-vac}$ mechanism.

## ACKNOWLEDGMENTS

This work was supported by the Assistant Secretary for Energy Efficiency and Renewable Energy, Vehicle Technologies Office, under the Applied Battery Materials Program, of the U.S. Department of Energy under contract no. DE-AC02-05CH11231. The computational analysis was performed using computational resources sponsored by the Department of Energy's Office of Energy Efficiency and Renewable Energy and located at the National Renewable Energy Laboratory as well as computational resources provided by Extreme Science and Engineering Discovery Environment (XSEDE), supported by the National Science Foundation grant number ACI1053575, and the National Energy Research Scientific Computing Center (NERSC), a DOE Office of Science User Facility supported by the Office of Science and the U.S. Department of Energy under contract no. DE-AC02-05CH11231. ZJ, LBL, and TC acknowledge financial support from the NSF Graduate Research Fellowship Program (GRFP) under contract no. DGE 1752814, DGE 1752814, and DGE 1106400, respectively. Any opinions, findings, conclusions, or recommendations expressed in this material are those of the author(s) and do not necessarily reflect the views of the National Science Foundation.

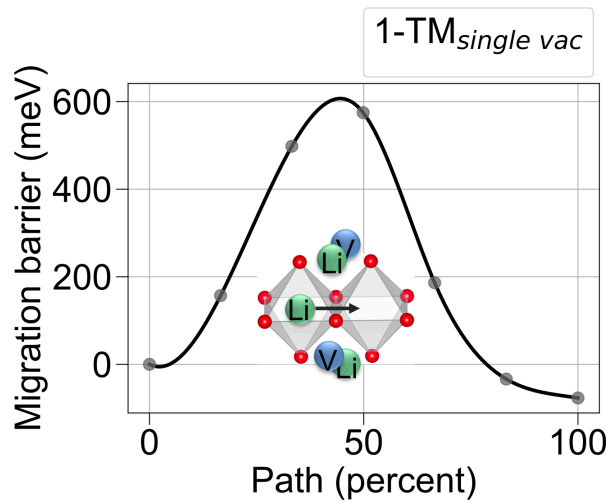

**Figure S1:** Energy (meV) along the Li migration pathway (%) for the octahedral-to-octahedral mechanism in Li$_2$VO$_3$ is examined when the lower and upper tetrahedral sites face-share with Li, V and one vacancy. Since the migration barrier exceeds 300 meV, it is safe to conclude that Li does not pass through the 1-TM single vacancy path. In addition, the intermediate tetrahedral site does not represent a local minimum in this case. Hence, we did not consider the 1-TM single vacancy hop in our KMC simulation.

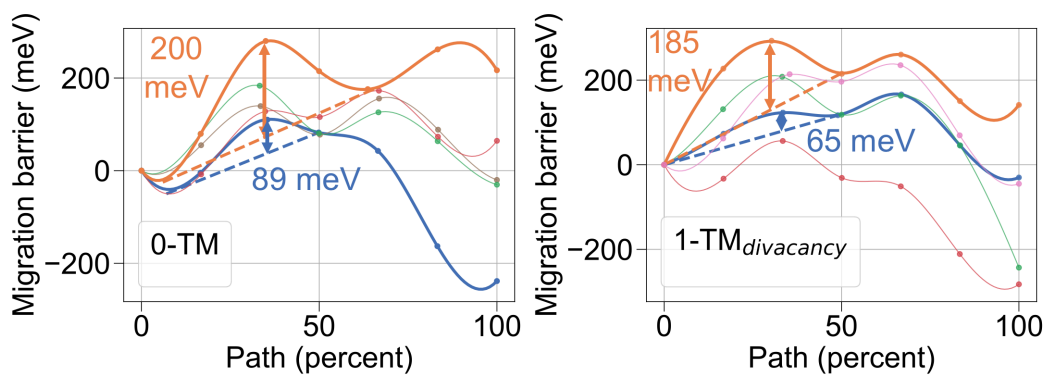

**Figure S2:** Energy (meV) along the Li migration pathway (%) for the octahedral-tetrahedral-octahedral mechanism in Li$_x$VO$_3$ (1 ≤ x ≤ 2): (left) when the tetrahedral site does not share a face with any V and (right) when it shares a face with one V and two octahedral vacancies. The minimum and maximum KRA values are indicated in each case.

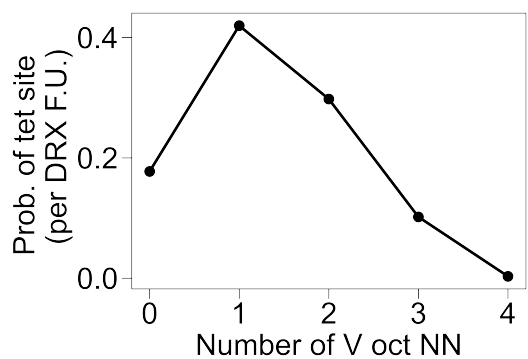

**Figure S3:** Probability of any tetrahedral site per DRX formula unit (F.U.) with a certain number of nearest-neighbor face-sharing V.